\documentstyle[12pt]{article}
\advance\textheight by \topskip
\begin{document}
\begin{flushright}
SU-ITP-95-26\\
hep-th/9511115\\
November 15, 1995\\
\end{flushright}
\vspace{-0.2cm}
\begin{center}
\baselineskip=16pt

{\Large\bf   SUPERSYMMETRIC  BALANCE OF FORCES  \\
\vskip 0.6 cm
AND   CONDENSATION OF BPS  STATES}\\

\

{\bf Renata Kallosh}\footnote {E-mail:
kallosh@physics.stanford.edu} {\bf ~ and~~ Andrei Linde}\footnote{E-mail:
linde@physics.stanford.edu}
 \vskip 0.05cm
Physics Department, Stanford University, Stanford   CA 94305\\
\vskip 0.7 cm

\end{center}
\vskip 1 cm
\centerline{\bf ABSTRACT}
\begin{quotation}

Until now all known static multi black hole solutions described BPS states with
charges of the same sign.  Such solutions could not be related to flat
directions in the space of BPS states: The total number of such states could
not spontaneously increase because of the charge conservation.

We show that there exist static   BPS configurations which remain in
equilibrium even if they consist of states with opposite electric (or magnetic)
charges from vector multiplets. This is possible because of the exact
cancellation between the Coulomb and scalar forces. In particular, in the
theories with N=4 or N=2 supersymmetry there exist stable massless multi center
configurations with vanishing total charge. Since such configurations have
vanishing energy and charge independently of their number, they can be
associated with flat directions in the space of all possible BPS states.   For
N=2 case this   provides a realization of the idea that BPS condensates could
relate to each other different vacua of the string theory.

\end{quotation}
\newpage
\baselineskip=16pt

\section{Introduction}

Recently it was found that there exist supersymmetric BPS states with vanishing
ADM mass. They include the solutions  with one half of the N=4
\cite{Klaus2,K,KL, CY}
or  N=2  \cite{FKS} supersymmetry unbroken. As usual for supersymmetric
configurations,  the multi-center solutions are also available \cite{K}. Those
solutions have various extraordinary features; in particular they do not
represent an  extreme limit
of any known non-extreme black holes. When considered in the four-dimensional
canonical frame, they are singular and the singularity has a universal
repulsive character. This is why it  is not appropriate to call them "black"
holes,  and  we   called them "white" holes, or repulsons \cite{KL}. In what
follows we will try to avoid attributing any color to the massless
supersymmetric configurations under study to minimize the abuse of terminology
\cite{Town},
 but we will keep calling them holes, or massless BPS states.

The existence of such solutions may look surprising \cite{Town}. However  one
should take into account several considerations:

i) The existence of such states in exact quantum theory is expected according
to conjectures of  dualities and enhanced symmetries \cite{HullT,W,Strom}.

ii)  BPS states of extended supersymmetry, if they exist as solutions of
classical theory,  may remain exact
in quantum theory, at least for N=4 case according to supersymmetric
non-renormalization theorems.

iii) If they would not be found at least one of the standard lores   i) or ii)
would be incorrect.

With  all this in view one may take the following attitude. Supersymmetric
black$\&$white holes
may have naked singularities and a better description of physical states of
supersymmetric gravity is desirable. However, so far they served well by
explicitly realizing various mass formulas for BPS states and mass  and charge
relations. The present investigation will show again that the new type of
mass-charge relations which are believed to remain exact in quantum theory can
be discovered when new soliton configurations are found.

The purpose of this note is to  exhibit some dramatic differences between new,
massless  supersymmetric configurations and the ones which are known as extreme
massive multi black hole
solutions. The difference is due to the fact that massless solutions  exist  in
case that we have more than one supersymmetric multiplet involved in the
solution, let us say $1+n$ multiplets. This leads to the existence of  $1+n$
independent supersymmetric balance of forces conditions.  The analysis of these
 conditions  shows that massless BPS states may consist   of the hole-anti-hole
configurations with the vanishing net charge which are in equilibrium.  We will
indeed find the multi-center BPS
configurations with the vanishing total mass and charge. The mass of every hole
 vanishes.   However,  the sign of the   charges with respect to   vector
fields other than graviphoton  may alternate from one hole to another. One may
argue, therefore that  the massless   holes may form a condensate. The argument
for this is simple: if  there is a state with $s$ holes which form a massless
neutral set and the other state with $s+t$\, holes which are also massless and
neutral, it will not cost any energy  to produce them, and no charge
conservation will be violated when the system changes from an
$s$-hole state to $(s+t)$-hole state.   In particular,   the total number of
massless neutral  clusters of $s$ holes may take any value. The effective
theory will have a non-trivial ground state with the massless mode
corresponding to the condensation of the neutral massless BPS configurations.

We would like to explain here under which conditions the massless neutral
solutions
can be found. Four-dimensional theories with one gravitational multiplet do not
have
non-trivial massless BPS configurations.
Indeed, with one multiplet of N=4 supersymmetry any solution is characterized
by the mass, which due to supersymmetry is related to the graviphoton charge
and to the dilaton charge. When this parameter vanishes the solution becomes
trivial.
With one gravitational multiplet of N=2 supersymmetry we have the relation
between the mass and the graviphoton charge. When the mass goes to zero, the
charge goes to zero.

In N=8 case   only the  gravitational multiplet exists. It is characterized by
one parameter  and when the mass goes to zero, only trivial solutions can be
expected. Unless more than one half of N=8 supersymmetry is broken, one cannot
expect any interesting massless solutions.

The situation changes  for N=4 theory since one can consider the interaction of
N=4 supergravity with any number of N=4 vector multiplets. In particular, one
can consider the theory with 22\, N=4 vector multiplets. Together with
6 graviphotons in the gravitational multiplet they form the original 28 vectors
 which previously were all in one gravitational multiplet of N=8 supergravity.
However, when the symmetry between 6 and 22 is broken by placing them into
different multiplets, we may expect that there exists a configuration with 6
graviphoton charges
vanishing but with 22  non-vanishing vector multiplet charges. This is indeed
the case \cite{Klaus2,K,KL, CY}.

If even more supersymmetries are broken and we are looking for configurations
with one half of N=2 supersymmetry unbroken, we may consider one gravitational
multiplet of N=2 supergravity interacting with some number of vector multiplets
of N=2 supersymmetry. During this step of breaking supersymmetry, the dilaton
which was in one multiplet with graviton is now in a different multiplet, in a
vector one. The dilaton charge, as well as the charges of the vector fields in
the vector multiplets, are not related to the graviphoton charge anymore.
Therefore one may have expected that
when supersymmetric solutions in this theory will be found, that they will
remain non-trivial when the mass of the configuration will tend to zero, since
the breaking of supersymmetry has made the charges of some vectors and scalars
not related to that of the graviphoton anymore. This is indeed the case: there
is a rich variety of massless solutions  in N=2 theory \cite{FKS}.

 \section{Supersymmetric balance of forces for massive extreme black holes:
gravitational multiplet solutions}

Consider an example of supersymmetric balance of forces between two $a=1$
stringy black holes of Gibbons-Maeda-Gurfinkel-Horowitz-Strominger.  These
solutions are supersymmetric when embedded into pure N=4 supergravity without
vector multiplets.
Any extreme solution in this class with one half of N=4 supersymmetry unbroken
saturate the following supersymmetric bound \cite{US} (we consider the electric
solutions first):
\begin{equation}
M^2 +( \Sigma_{\rm dil})^2  - (Q_{\rm gr})^2 = 0 \ .
\end{equation}
The mass $M$ is related to the graviphoton charge $Q_{\rm gr}$ and to the
dilaton charge $ \Sigma_{\rm dil}$ as follows:
\begin{equation}\label{BPSEL}
M= \frac {|Q_{\rm gr}|}{\sqrt{2}} =- \Sigma_{\rm dil}  \ .
\end{equation}
If there are two supersymmetric black holes, each of them has to saturate the
same type of bounds:
\begin{eqnarray}
m_1^2 + \sigma_1^2 - q_1^2  &=& 0 \ ,\nonumber\\
\nonumber\\
m_2^2 + \sigma_2^2  - q_2^2 &=& 0 \ .
\end{eqnarray}
For each multiplet we have the same set of relations between charges in every
hole.
\begin{eqnarray}
\label{multi}
 m_1 &=& -\sigma_1 = \frac
{|q_1|}{\sqrt{2}} \ , \nonumber\\
\nonumber\\
m_2 &=& -\sigma_2 = \frac {|q_2|}{\sqrt{2}}\ .
\end{eqnarray}
We can study the forces between two such black holes and find whether the holes
are in equilibrium.  Consider Newtonian,
Coulomb and
dilatonic forces between two distant objects of masses
and charges
$(m_1, q_1, \sigma_1)$ and $(m_2, q_2, \sigma_2)$:
\begin{equation}\label{bal}
F_{12} = - \frac{m_1 m_2}{r_{12}^2} +
\frac{q_1 q_2}{r_{12}^2}  - \frac{\sigma_1 \sigma_2}{r_{12}^2} \ .
\end{equation}
The dilatonic force is attractive for charges of the same sign and
repulsive for charges of opposite sign. For our configurations {\it the dilaton
charge does not change the sign when the electric charge does}. Therefore if
the electric charges of the two black holes are of the same sign we get
\begin{equation}\label{bal1}
F_{12} = - \frac{m_1 m_2}{r_{12}^2} +
\frac{|q_1| |q_2|}{r_{12}^2}  - \frac{\sigma_1 \sigma_2}{r_{12}^2} \ .
\end{equation}
According to eq. (\ref{multi}),
\begin{equation}\label{bal2}
F_{12} = - \frac{m_1 m_2}{r_{12}^2} (1 -2+1) =0
 \ .
\end{equation}
Thus we see  that $F_{12}$ vanishes and we conclude that two holes with the
same sign of the electric charge  are in  equilibrium with each other, since
the
attractive
force between them due to gravity and dilaton force is cancelled by Coulomb
repulsion of two equal sign charges.

Now consider the situation when the electric charges have opposite sign. All
three forces are attractive this time:
\begin{equation}\label{bal3}
F_{12} = - \frac{m_1 m_2}{r_{12}^2} (1 +2+1) = \frac{4 m_1 m_2}{r_{12}^2}\neq 0
 \ .
\end{equation}
This configuration is unstable since there is no balance of forces. Instead of
performing   this balance of forces analysis one could simply look into
explicit form of the two-hole solution \cite{US}. There is no static solution
available   for two holes  with positive masses and with opposite electric
charges.

 For a magnetic BPS state  with the mass $M$, the magnetic graviphoton charge
$P_{\rm gr}$ and the dilaton charge $ \Sigma_{\rm dil}$ the relation between
charges is similar to (\ref{BPSEL}):
\begin{equation}
M= \frac {|P_{\rm gr}|}{\sqrt{2}} = \Sigma_{\rm dil}  \ .
\end{equation}
The conclusions of the balance of force analysis would be the same: one can get
two holes in equilibrium under the condition that they have magnetic charges of
 the same sign.

Thus one could have concluded that two supersymmetric   black holes may be in
equilibrium only when their charges have the same  sign. Therefore the sum of
all masses of BPS states in equilibrium and the absolute value of their
graviphoton charges   is always positive:
\begin{equation}
\sum_{a=1}^{n} m_a > 0 \ ,  \qquad |\sum_{a=1}^{n} (q_{\rm gr}) _a| > 0  \ .
\end{equation}

This is were the massless configurations brought a new surprise.

 \section{Supersymmetric balance of forces: gravitational and vector multiplet
solutions}

As explained above we have explicit solutions with vanishing ADM mass only in
case of few multiplets, one gravitational and $n$ vector multiplets. In case of
N=4 supersymmetry our solutions saturate $1+n$ independent bounds, one for each
multiplet.
\begin{eqnarray}
&&M^2 +( \Sigma_{\rm dil})^2  - (Q_{\rm gr})^2 = 0 \ , \\
&& ( \Sigma^I_{\rm vec})^2  - (Q^I_{\rm vec})^2= 0 \ , \qquad  I=1, \dots , n.
\end{eqnarray}
The first bound relates three type of charges: the mass, the dilaton charge and
the electric charge of the graviphoton in the gravitational multiplet. Every
other bound relates the modulus charge $\Sigma_{\rm vec}$ with the vector
charge
$Q_{\rm vec}$ inside each vector multiplet.

The nature of the second type of supersymmetric bounds turned our to be very
different from what may have been expected.  The important property of the
supersymmetric solutions \cite{Klaus2,K,KL, CY,FKS}  is the following: the sign
of the charge of the modulus field  $\Sigma^I_{\rm vec}$ is correlated with the
sign of the charge of the electric field of the same vector multiplet.
\begin{equation}\label{chargerel}
\Sigma^I_{\rm vec}  = Q^I_{\rm vec} \ .
\end{equation}
When the sign of the vector charge changes, the sign of the scalar charge also
changes. Therefore it is possible to consider  the situation when there are two
massive black holes, saturating all  bounds.  The picture described above for
one gravitational multiplet remains intact: the graviphoton charge must always
be of the  same sign for two holes. However, the vector multiplet charges do
not have to be constrained that way.
Each black hole has to saturate its own vector multiplet bound.
\begin{equation}
( \sigma^I_{\rm vec})^2_{1}   - (q^I_{\rm vec})^2_{1} = 0 \ , \qquad
( \sigma^I_{\rm vec})^2_{1}   - (q^I_{\rm vec})^2_{2} = 0  \ .
\end{equation}
Given that the vector charge of every hole equals its scalar charge,
\begin{equation}
(\sigma^I_{\rm vec})_{1}  = (q^I_{\rm vec})_{1} \ ,   \qquad (\sigma^I_{\rm
vec})_{2}  = (q^I_{\rm vec})_{2} \ ,
\end{equation}
all bounds are saturated and there is  no restriction on the relative sign of
vector  charges of two holes.
There is always  the balance of forces:
whether we have the Coulomb attraction for  opposite  sign electric charges or
Coulomb repulsion for same sign electric charges, we get the same picture from
modulus field force. Since the sign of the scalar charge is correlated with the
sign of the vector charge the corresponding part of the force between two holes
vanishes
\begin{equation}\label{bal4}
F_{12}^{\rm vec} = \frac{ (q^I_{\rm vec})_{1} (q^I_{\rm vec})_{2}}{r_{12}^2}  -
\frac{(\sigma^I_{\rm vec})_{ 1} (\sigma^I_{\rm vec})_{ 2}}{r_{12}^2}=0 \ .
\end{equation}
Thus we have found that for the massive extreme black holes with graviphoton
charges as well as with vector multiplet charges there are two possible
configurations in equilibrium: The sign of graviphoton charge for two holes is
always the same, but the sign of the vector multiplet charge may or may not
alternate between two holes. One may have either
\begin{equation}
(q_{\rm gr})_1  (q_{\rm gr})_2 > 0 \ , \qquad (q^I_{\rm vec})_{1} (q^I_{\rm
vec})_{2} >0 \ , \end{equation}
or
\begin{equation}
(q_{\rm gr})_1  (q_{\rm gr})_2 > 0 \ , \qquad (q^I_{\rm vec})_{1} (q^I_{\rm
vec})_{2} <0 \ . \end{equation}

In N=2 case we have found extreme black hole solutions \cite{FKS} which
saturate the gravitational bound as well as the vector multiplet bound:
\begin{eqnarray}
&&M^2   - (Q_{\rm gr})^2 = 0 \ , \\
&& ( \Sigma^I_{\rm vec})^2  - (Q^I_{\rm vec})^2 = 0 \ , \qquad  I=1, \dots , n.
\end{eqnarray}
The difference from N=4 case is in the structure of the gravitational bound
which does not include the dilaton charge anymore. It is saturated when $M=
|Q_{\rm gr}|$.
The dilaton (in heterotic theory) is now placed in one of the vector multiplets
and as such, behave as any other scalar of the vector multiplets when the
vector charge changes the sign. Alternatively, from type II string theory point
of view, the dilaton is
in one of the hypermultiplets. Apart from this there is no difference in the
picture of balance of forces, as discussed above: the graviphoton charge splits
into the same sign charges whereas
the vector multiplet charge may split into same sign or opposite sign charges.

The crucial feature of the configuration with the vanishing mass, dilaton
charge and graviphoton charge is the fact that the gravitational supersymmetric
bound is saturated trivially for each hole. There is no gravitational, dilaton
and electric graviton force to be compensated, all of them vanish in order
${1\over r^2_{12}}$.
Thus the conclusion from the balance of forces study is: the configurations
with the vanishing  total mass  and all charges
may exist.
\begin{equation}
M= \sum_{a=1}^{n} m_a = 0 \ , \qquad
Q_{\rm gr}= \sum_{a=1}^{n} (q_{\rm gr}) _a=0 \ , \qquad
 Q^I _{\rm vec }=\sum_{a=1}^{n}
(q_{\rm vec }^I) _a = 0  \ .
\end{equation}
The massless solutions with  the vanishing
 vector multiplet charge  $Q^I _{\rm vec }=0$ in all gauge groups  were not
known to exist before. The  mechanism of the black hole condensation proposed
by Greene, Morrison and Strominger \cite{GMS}
(GMS mechanism) is based on  the assumption that  such configurations exist.
Our analysis of the balance of forces proves that they may exist. In what
follows we
will describe them.

\section{Hole-anti-hole solution in N=4 theory}

We would like to consider first the two-hole and $2s$-hole supersymmetric
solution which may have an interpretation of the particle-antiparticle state.
The class of massive black holes with non-vanishing graviphoton charge will not
allow us to find such, as explained above. However, if we  use directly the
massless solutions,  those with two opposite charges can be found.
Starting with the Maharana-Schwarz-Sen action for toroidally compactified
heterotic string theory  we can get a two-center supersymmetric
massless solution by slightly generalizing the construction presented in
\cite{BK}.

The  generic pure magnetic $(6,22)$ symmetric solution is described completely
in terms of  magnetic potentials  $\vec \chi$:
\begin{equation}
\vec{\chi}(x) = \left( \begin{array}{c} \vec{\chi}_{\rm vec} (x)\\
\vec{\chi}_{\rm gr}(x)
             \end{array} \right)
	     	    \ , \qquad
 \partial_i  \partial_i  \vec{\chi}(x) =0 \ .
\label{magn}\end{equation}
 The 28-dimensional  harmonic  $O(6,22)$-vector  $\vec \chi$ consists of the
22-dimensional vector
$\vec{\chi}_{\rm vec}$ , or
$\chi ^I,\, I=1,\dots 22$, describing the vector multiplets  and of the
6-dimensional vector $\vec{\chi}_{\rm gr}(x)$,
 or $\chi^\alpha ,\,  \alpha=1,\dots 6$, describing the gravitational
multiplet. The  metric, the dilaton, the moduli
fields ${\cal M}$ and the magnetic field $ \vec{H}_i =  {1\over 2}
\epsilon_{ijk} \vec {F}_{jk} $ are
\begin{equation} \label{monopole}
\begin{array}{ccc}
ds^2_{\rm can}= - e^{2 U} dt^2 + e^{-2 U} d\vec{x}^2\ , &&  e^{-4U} = 2 \,
\left ( (\chi^\alpha)^2 - (\chi^I)^2\right)
                                              = e^{4\phi} \ , \\
\nonumber\\
\nonumber \\
{\cal M}= {\bf 1}_{28} + 4 e^{4U} \left( \begin{array} {cc}  \chi^I  \chi^J  &
\chi^ I  \chi^\beta
\\
 \chi^\alpha  \chi^J    &     \xi \;  \chi^\alpha  \chi^\beta \end{array}
\right) , & &  \vec{H}_i =  \partial_i \vec{\chi}\ ,
 \end{array}
\label{multisol}\end{equation}
where $ \xi \equiv  (\chi^I)^2 /  (\chi^m)^2$.

Consider  the simplest case of asymptotically flat geometry and vanishing at
infinity scalar fields.  The
massless  solution with $a=2s$ holes  and vanishing net charge for each of the
28 gauge groups is given by\footnote {This solution with same sign charges was
found in \cite{BK}, however it
remained unnoticed that alternating signs are also possible.}
\begin{equation}
\chi^I  = \sum_{a=1}^{a=2s}  \frac{q ^I{}_a }{|\vec x- \vec x_a|} \  , \qquad
\chi^\alpha = {1\over \sqrt 2} n^\alpha \ , \qquad  (n^\alpha)^2 =1 \ ,
\end{equation}
\begin{equation}
Q^I \equiv   \sum_a q ^I{}_a = 0 , \qquad  m_a=0 \ , \qquad M= \sum m_a =0 \ .
\end{equation}
The graviphoton charge vanishes for each hole  since each of them is massless.
However, the
vector multiplet charge of each individual hole does not vanish, its sign
alternates and only the total sum over all holes vanishes.
For example, the simplest monopole-anti-monopole  solution has two holes of the
opposite charge,
 \begin{equation}
\chi^I  =   \frac{q ^I }{|\vec x- \vec  x_1|}   -  \frac{q ^I }{|\vec x-\vec
x_2|} \  , \qquad
\chi^\alpha = {1\over \sqrt 2} n^\alpha \ , \qquad  (n^\alpha)^2 =1 \ ,
\end{equation}
\begin{equation}
Q_I \equiv   q ^I{}_1 +  q ^I{}_2 = 0 , \qquad  m_1=m_2 = 0 \ , \qquad M=  m_1
+ m_2 =0 \ .
\end{equation}
To confirm our balance of force condition analysis we must check that the
moduli field charges indeed compensate the Coulomb forces between the
monopole-anti-monopole pair. For this purpose we will write down the total
solution, corresponding to the potential of the pair above.
We have for the metric,   the dilaton and the magnetic fields:
\begin{equation}
  e^{-4U} = \, 1 - 2 \Bigl( \frac{q ^I }{|\vec x-\vec x_1|}   -  \frac{q ^I
}{|\vec x-\vec x_2|}\Bigr)^2   = e^{4\phi} \ ,  \qquad H_i ^I = { (x-x_2)^i q^I
\over |\vec x- \vec x_2|^3}-
{ (x-x_1)^i q^I \over |\vec x-\vec x_1|^3} \ ,
\end{equation}
and there are no graviphoton magnetic fields, $H_i^\alpha=0 $. The moduli
fields are given in eq.
(\ref{multisol}). We are particularly interested here in the moduli field
charges. Therefore we will write down
explicitly only the terms  required for defining the scalar field charges.
Those are \begin{equation}
{\cal M}= {\bf 1}_{28} + 2\sqrt2 \, \pmatrix{
 0&    n^\beta\left(  \frac{q ^I }{|\vec x- \vec  x_1|}   -   \frac{q ^I
}{|\vec x-\vec x_2|}  \right)  \cr
 \cr
 n^\alpha \left( \frac{q ^J }{|\vec x- \vec  x_1|}   -  \frac{q ^J }{|\vec
x-\vec x_2|} \right) & 0 \cr
} + \dots
\end{equation}
Here $\dots $ stays for terms which will not contribute to scalar charges of
each monopole. To find  the charges of the ${\cal M}^{I\beta }$ and  ${\cal
M}^{\alpha J}$ we will take the following two limits. First, we choose $\vec
x_1=0$, i.e. we place the first hole in the center of coordinates.
The distance between two holes is given by the vector $\vec l$. The scalar
${\cal M}^{\alpha J }$ becomes
\begin{equation}
{\cal M}^{\alpha J }= 2 \sqrt 2 n^\alpha \left( \frac{q ^J }{|\vec x|}   -
\frac{q ^J }{|\vec x-\vec l |} \right) \ .
\end{equation}
The scalar charge of the first hole is defined as follows. We remove the second
hole far away,
i.e.
we consider $l \rightarrow \infty$. In this limit
\begin{equation}
{1\over 2 \sqrt 2} ({\cal M}^{\alpha J })_1 \rightarrow     \frac{ n^\alpha q
^J }{|\vec x|}    \equiv
{(\Sigma ^{\alpha J } )_1 \over  | \vec x|} \ .
\end{equation}
This defines the scalar charge of the first hole. In our case this leads  to
\begin{equation}
(\Sigma ^{\alpha J })_1 =  n^\alpha q ^J =  n^\alpha (q ^J)_1 \ .
\end{equation}
To find the scalar charge of the second hole we place it  in the beginning of
coordinates, remove the first  hole far away and consider  the $(\Sigma
^{\alpha J } )_2 / | \vec x|$ term. We get
\begin{equation}
(\Sigma ^{\alpha J })_2 =- \, n^\alpha q ^J =  n^\alpha (q ^J)_2 \ .
\end{equation}
Thus we have confirmed the balance of force analysis: the scalar charge of the
vector multiplet is sensitive to the sign of the magnetic charge of the same
multiplet. Therefore the supersymmetric positivity bound permits massless holes
with opposite charges to be in equilibrium.

 The reader familiar with Sen's spherically symmetric extreme massive
supersymmetric black holes \cite{Sen}  may easily verify that the dilaton
charge of the electric
(magnetic)  solution is not sensitive to the sign of the electric (magnetic)
graviphoton charge $Q_R$, whereas the charge of the modulus ${\cal M}$ does
change when the sign of the charge of the vector in the vector multiplet  $Q_L$
changes.
This may give an additional explanation of why  the BPS solutions with half of
unbroken supersymmetry include massless neutral  configurations with arbitrary
number of centers.

We should emphasize, that the solutions discussed above are obtained by solving
exact nonlinear equations. The balance of force analysis, which describes
asymptotic behavior of forces between the two holes, is necessary only to
interpret our exact solutions and to explain why they are consistent despite
describing configurations with opposite charges.

Black hole multiplets with unbroken N=4 supersymmetry form  the vector
multiplets of N=4 supersymmetry. It is likely that the dynamics of such
multiplets  will show the condensation of the massless states. Indeed, it does
not cost any energy to create as many pairs  of massless oppositely charged BPS
states as one wishes, and it does not violate charge conservation since each of
 these pairs is electrically and magnetically neutral. We will consider below
the case of N=2 theory which is simpler from the point of view of the effective
theory and which also has the massless neutral BPS states.

\section{Condensation of Massless Holes in N=2 theory}
The GMS mechanism  of black hole condensation in N=2 theory is the following
\cite{GMS}. One starts with the system of 16 massless black holes of unbroken
N=2 supersymmetry. The low energy theory should contain 15 $U(1)$ gauge groups.
The total charge of the system of 16 holes in each of the 15 gauge groups has
to vanish,
\begin{equation}
Q^I \equiv \sum_{a=1}^{16} (q^I)_a =0 \ .
\label{net}\end{equation}
A specific example in  \cite{GMS} was to have the first hole with the charge +1
in the first group,
the second one with the charge +1 in the second group, the third one with the
charge +1 in the third group, etc. The 16-th hole, however, had to be
negatively charged in all 15 groups.
The problem which was not clearly resolved there was whether massless black
holes can actually exist and whether they can be in equilibrium despite the
Coulomb attraction between objects with opposite charges.

The  Calabi-Yau manifold in question was described in a way that it is
difficult to find what kind of a Kahler manifold corresponds to it in the low
energy four-dimensional theory. Therefore, we will not consider the study below
as the one describing a particular Calabi-Yau manifold. However, we will  give
an example of N=2 black holes which have the massless limit and
have the set of 16 holes with the charges satisfying the constraint (\ref{net})
in equilibrium.
The simplest case is to use the example of
 ${ SU(1,15 ) \over SU(15)} $~   N=2 black holes \cite{FKS}.

The prepotential and the Kahler potential are
\begin{equation}\label{36}
F(X^0,X^I) = (X^0)^2- (X^I)^2 \ , \qquad  e^{- K(Z,\bar Z)} = 1-|Z^I|^2 \ .
\end{equation}
We choose 15 harmonic scalars  $Z^I={X^I\over X^0}$  to vanish at infinity and
to describe a 16-hole massless configuration:
\begin{equation}\label{15}
 Z^I   =  \sum _{a= 1}^{16} {q^I{}_a \over |\vec x - \vec x_a| }\, \ , \qquad
\sum _{a= 1}^{16} q^I{}_a=0 \ ,
\end{equation}

\begin{equation}\label{39}
ds^2 = (1-|Z|^2)^{-1}\, dt^2 -
(1-|Z|^2)\,
d\vec x^2 \ .
\end{equation}
The magnetic charges of 15 vector fields are given by $q^I{}_a$. In a more
detailed form and assigning only $\pm 1$ charges we have
\begin{eqnarray}\label{16}
Z^1 &=& {1 \over |\vec x - \vec x_1| } - {1 \over |\vec x - \vec x_{16}| }\ ,
\\
Z^2 &=& {1 \over |\vec x - \vec x_2| } - {1 \over |\vec x - \vec x_{16}| }\ ,
\\
& & \dots \\
Z^{15} &=& {1 \over |\vec x - \vec x_{15}| }- {1 \over |\vec x - \vec x_{16}|
}\ .
\end{eqnarray}
The total ADM mass, the total magnetic charge in each of the 15 gauge groups,
as well as the total scalar charge in every gauge group vanish. Still we have a
rather non-trivial configuration. If we take only one hole, for example the
first one, remove all other 15 far away, we will find that
it has a non-vanishing positive magnetic as well as the scalar charge. The same
with the second and other 13 holes. If we  finally consider the 16-th one, we
would find that when the other 15 holes are far away, this one happens to be
negatively charged (both in magnetic and scalar charges) in all 15 gauge
groups. The system of these 16 holes is in equilibrium.

It is natural therefore to move to an alternative picture: associate with each
hole a massless hypermultiplet   as suggested in \cite{GMS}. Each
hypermultiplet  contains two  charged
complex scalars
 $h^{a\alpha}$ where $\alpha = 1,2$ is the global $SU(2)_R$ index of
N=2 representation. This gives a total of $32$ complex scalar fields
 The potential describing the interactions between these holes in agreement
with N=2
supersymmetry is given by
\begin{equation}\label{pot}
V\sim \sum_{I,J=1}^{15} M_{IJ}D ^{\alpha \beta  I}D_{\alpha \beta}{}^{ J} \ ,
\end{equation}
where $M_{IJ}$
is a positive definite matrix  and
\begin{equation}
D^{\alpha \beta  I}  = \sum_{a=1}^{16} q^I{}_a ( h^{* a \alpha} h^{a\beta} +
 h^{* a \beta} h^{a \alpha}) \ .
\end{equation}
A remarkable feature of this potential is the existence of  flat directions
along which the bilinear combinations of scalar fields $D^{\alpha \beta I}$
vanish,
\begin{equation}
D^{\alpha \beta I} = 0 \ .
\label{flat}\end{equation}
This gives $45$ real constraints on  $32$ complex fields. In addition there
are $15$ gauge transformations which rotate the fields, leaving 4 real
vacuum parameters. Up to a
gauge transformation the general solution of (\ref{flat}) is defined by a
complex two-vector $v$.
\begin{equation}
h^{a\alpha} = v^{\alpha}\qquad \hbox{for all $a$} \ .
\end{equation}
 Moving along the flat direction, the  holes condense and their moduli space
is parametrized by a single hypermultiplet. The point $v = 0$ was associated in
\cite{GMS} with
the conifold point in the space of quintics (at which all $16$ cycles vanish).
Moving away from this point along the flat direction corresponds to giving  a
vev's to
the charged hypermultiplets which break all $15$ U(1)'s. Thus  a second branch
of the moduli space corresponding to a
charged black hole condensate was discovered in \cite{GMS}. This branch has
$101-15=86$ massless vector multiplets, and $2+1=3$ massless hypermultiplets.

We would like to stress here that the total picture depends crucially on the
fact that $D^{\alpha \beta I} = 0
$  when the condensate ansatz is introduced into the expression for the flat
direction
condition.  The reason it works is that
\begin{equation}
D^{\alpha \beta I}|_{ h^{a \alpha} = v^{\alpha}} = ( v^{*  \alpha} v^{\beta} +
 v^{*  \beta} v^{ \alpha})  \sum_{a=1}^{16} q^I{}_a =   ( v^{*  \alpha}
v^{\beta} +
 v^{*  \beta} v^{ \alpha})\; Q^I =0 \ .
\end{equation}
Thus it is clear that if the total charge $Q_I$ of all holes in every gauge
group  would not vanish, we would be unable to have a condensate with
$v^{\alpha} \neq 0$.
But now that we know that such 16 holes may exist in equilibrium the total
picture of condensation of massless holes with vanishing total charge looks
much more plausible.

\section{Massless Mode as a Goldstone Boson}

Being supported by the existence of exact massless neutral multi-center
solutions, we may study the general case of   spontaneous symmetry breaking
which brings one theory with $h_{21}$ massless vector
 multiplets and $ h_{11}$ massless hypermultiplets into the  phase with
$h'_{21}$ massless vector multiplets and    $ h'_{11}$ massless
hypermultiplets.

Let us first try to understand  if there is any  magic here about the numbers
15 and 16 in the example above.
If we would start with $n$ gauge groups instead of 15 and $n+1$ holes
(hypermultiplets) we would easily construct an\, $n+1$-hole solution using the
example in \cite{FKS} with  ${ SU(1, n ) \over SU(n)} $.  This would be the
generalization of eqs. (\ref{15}), (\ref{16}). Thus we can consider the case of
 $n$ gauge groups and $n+1$-hole solution with the total mass and charges in
every gauge group vanishing.
However, what will happen with the counting above which worked well for 15,16
case? It actually works in the general case as well. We start with   $4 \times
(n+1)$ real scalars (for $n+1$ hypermultiplets). We have to impose
 $3\times n$ conditions for a direction to be flat. In addition we can use $n$
gauge transformations. Thus we are left with 4 real vacuum parameters
$v^\alpha$ as before, since previously we had
\begin{equation}
[4 \times (15+1)] - [3\times 15] -[15] = 4 \ ,
\end{equation}
and now we have
\begin{equation}
[4 \times (n+1)] - [3\times n] -[n] = 4 \ .
\end{equation}
This would describe a condensate of a system of the original $n+1$ holes. The
moduli space of the condensed holes is parametrized as before by a single
hypermultiplet. The condensate breaks $n$ gauge groups this time.

Thus, suppose we start with $n$ gauge groups and choose the $n+1$-hole solution
with the total charge in all groups vanishing. The  distribution of the charges
has the same structure as before: the first hole has charge 1 in the first
group, etc, the $n$-th one has the charge $1$ in the $n$-th group. The last one
has the negative charge in all groups.  One can represent it as a charge matrix
with $n$ columns (for $n$ groups) and $n+1$ rows (for  $n+1$ holes)
\begin{equation}\label{config}
\left (\matrix{
{}~~1 & ~0 ~\cdots & 0 \cr
{}~~0 & ~1 ~ \cdots & 0 \cr
{}~~\vdots &~~\vdots ~~\ddots & \vdots \cr
{}~~0 & ~0 ~\cdots & 1 \cr
-1 &\hskip -.3 cm -1 ~\cdots &\hskip -.2 cm -1~ \cr
}\right )
\end{equation}
The counting above shows that again we get one massless mode and $n$ massive
ones. The relevant question to ask is: does this situation allow the
interpretation of the massless mode as a Goldstone boson?
For this purpose we may check the symmetries of the potential in the form in
which we first perform the shift of the fields. The form of the hyper-hole
condensate $\langle h^a\rangle  = v$ suggests the natural combinations of
fields,\footnote{One can choose  different
forms of the hyper-hole condensate, e.g.  $\langle h^1\rangle  = v,~~ \langle
h^2\rangle  = - v $, etc.  For all of this choices one can find the natural
combinations of fields in terms of which the theory near each of these  ground
states is described as the one with $\langle h^a\rangle  = v$. }
\begin{equation}
\phi^{\alpha I} \equiv  \sum_{a=1}^{n} q^I{}_a  h^{ a \alpha}\ , \qquad
G^\alpha  \equiv
h^{(n+1)\alpha}\ ,
\end{equation}
such that the ground state is defined by
\begin{equation}
\langle \phi^{\alpha I}\rangle =0\ , \qquad \langle G^\alpha \rangle = v^\alpha
\ .
\end{equation}
The potential in these variables is given by eq. (\ref{pot}) with
\begin{equation}
D^{\alpha \beta I} = \phi ^{*  \alpha I} (  \phi ^{  \beta I} + 2 G^\beta) +
 (  \phi ^{*  \beta I} + 2 G^{* \beta}) \phi ^{ \alpha I}\ .
\end{equation}
In terms of fields with vanishing vacuum expectation values
\begin{equation}
G^\alpha = \tilde {G}^\alpha + v^\alpha\ , \qquad
\langle\tilde {G}^\alpha\rangle =0\ ,
\end{equation}
 we have
\begin{equation}
D^{\alpha \beta I} = \phi ^{*  \alpha I} (  \phi^{  \beta I} + \tilde {G}^\beta
+ v^\beta) +
 (\phi ^{*  \beta I} + \tilde {G}^{* \beta} + v^{*\beta}) \phi ^{ \alpha I}\ .
\end{equation}
The relevant Goldstone-type continuous  global symmetry of this theory is a
$U(2)$  symmetry given by
\begin{equation}\label{sym}
\Delta \phi ^{  \alpha I} = \lambda^{\alpha \beta}  \phi^{  \beta I}\ , \qquad
\Delta  G^\alpha = \lambda^{\alpha \beta}
G^\beta\ ,  \qquad \lambda^\dagger = \lambda \ .
\end{equation}
This symmetry has exactly four parameters in agreement with the fact that one
massless hypermultiplet has four massless scalars.
This symmetry is broken spontaneously.
To prove that there is one massless Goldstone hypermultiplet (four scalars) in
such theory  one can use the standard procedure. The values of the variations
of the fields  at the ground state are
\begin{equation}
\langle\Delta \phi^{  \alpha I}\rangle  =  \lambda^{\alpha \beta}
\langle\phi^{  \beta I}\rangle  =0\ ,  \qquad \Delta \langle G^\alpha\rangle  =
\lambda^{\alpha \beta}
\langle G^\beta\rangle  =   \lambda^{\alpha \beta} v^\alpha  \ .
 \end{equation}
{}From the symmetry of the potential we get
\begin{equation}
{\partial  V \over \partial \phi^I} \Delta \phi^I  + {\partial  V \over
\partial G} \Delta G + {\partial  V \over \partial \phi^{I*}} \Delta \phi^{I*}
+ {\partial  V \over \partial G^*} \Delta G^* =0 \ .
\end{equation}
The second derivatives of this equation over $G^*$ and $\phi^I$ taken at the
ground state give the consistency conditions for the mass matrix:
\begin{eqnarray}
\Big\langle{\partial^2 V \over \partial G^\alpha  \partial G^{*\gamma
}}\Big\rangle  \lambda^{\alpha \beta} v^\beta     + \Big\langle{\partial^2  V
\over \partial G^{*\alpha}  \partial G^{*\gamma}}\Big\rangle ( \lambda^{\alpha
\beta} v)^* &=&0 \ , \nonumber\\
\nonumber\\
\Big\langle{\partial^2  V \over \partial G^\alpha  \partial \phi^{*\gamma  I}
}\Big\rangle   \lambda^{\alpha \beta}  v    + \Big\langle{\partial  V \over
\partial G^{* \alpha} \partial \phi^{* I \gamma}}\Big\rangle ( \lambda^{\alpha
\beta} v)^* &=&0 \ .
\end{eqnarray}
This tells us that there is one Goldstone massless hypermultiplet $G^\alpha$
with four massless scalars whose mass matrix is decoupled from the other $n$
massive hypermultiplets $\phi_I^\alpha$. This is  in agreement with the fact
that there is one four-dimensional spontaneously broken continuous symmetry
$U(2)$
(\ref{sym}).
The spontaneously generated mass term for $n$ hypermultiplets is
\begin{equation}
\sim [\phi ^{*  \alpha I} v^\beta +
 v^{*\beta} \phi ^{ \alpha I}] M_{IJ} [\phi_{  \alpha} ^{* J} v_\beta +
 v^{*}_{\beta} \phi_{  \alpha}^J] \ .
\end{equation}
If we would start with the neutral set of holes but with the different
distribution of charges over the holes we would have to follow a  slightly more
complicated procedure of diagonalizing the massive and massless modes. However
we will get again $n$ massive and one massless mode.

Thus we have found how to decrease  any number $ h_{21} - h_{21}'=n $ of
massless vector multiplets and increase the number of massless hypermultiplets
by one ($ h_{11}'-
h_{11}=1$). The rule was to create one neutral cluster of $n+1$ massless
charged holes with the configuration of charges described in (\ref{config}). If
one would like to have more than one,  $k= h_{11}'-
h_{11}$, massless hypermultiplet as a result of the hole condensation, this is
also  possible. One has to start with $k$ clusters of  neutral system of holes
of the type
described above. Each cluster has some number of columns $n_k$   and rows
$(n_k+1)$.
\begin{equation}\label{matrix}
\left (\matrix{
\left (\matrix{
{}~1 ~~ \cdots ~~0 \cr
{}~\vdots  ~~\ddots ~~ \vdots \cr
{}~0 ~~\cdots  ~~1 \cr
-1 ~\cdots  -1 \cr
}\right )
& \cdots & 0 \cr
\vdots&  \ddots & \vdots \cr
\cr
0 & \cdots &
\left (\matrix{
{}~~1 & ~0 ~\cdots & 0 \cr
{}~~0 & ~1 ~ \cdots & 0 \cr
{}~~\vdots &~~\vdots ~~\ddots & \vdots \cr
{}~~0 & ~0 ~\cdots & 1 \cr
-1 &\hskip -.3 cm -1 ~\cdots &\hskip -.2 cm -1~ \cr
}\right )
 \cr
}\right )
\end{equation}
The matrix of charges is a direct product of the matrices of the type given for
one neutral set. The total number of columns in this matrix equals the number
of vector multiplets $n= n_1+n_2+\dots
+n_k$ which become massive in the process of spontaneous breaking of symmetry.
The number of submatrices in  (\ref{matrix}) equals $k$, the number of
massless hypermultiplets.
If the hypermultiplets in each cluster interact only with the vector multiplets
 and hypermultiplets of its own group one can write down the potential with
$[U(2) ]^k$ global symmetry, which is spontaneously broken and as the result,
$k$ massless hypermultiplets describe the condensed state of the original
neutral $k$ clusters.
All of such multi-hole configurations are available.

\section{After Condensation}
After the condensate of one cluster of massless holes has been formed we end up
by a set of holes which are all massive except one. It is interesting to go
back from the effective action which was treating the holes as hypermultiplets
to the original action
of supergravity interacting with $n$ vector multiplets and identify the
configuration which describes the state of a system after the condensate has
been formed.   For simplicity consider the case of  two sets of 4 gauge groups
and again, the simplest for our purpose Kahler manifold  describing
 ${ SU(1,4+4 ) \over SU(4+4)} $~   N=2 black holes \cite{FKS}.  The first four
massless vector multiplets will be excited in the  initial configuration and
the second four in the final\footnote{ We need two sets of vector multiplets
since the fields in the first group will become massive when the condensate is
formed.}.
This will allow us to have an example of 5 holes with the required properties.
The original configuration of 5 massless holes is given by
\begin{equation}
m_1 = m_2=m_3=m_4=m_5=0\ , \qquad Z^1_\infty =
Z^2_\infty=Z^3_\infty=Z^4_\infty=0 \ .
\end{equation}
with the charge matrix
\begin{equation}\label{1}
\left (\matrix{
{}~1 & ~0 & ~0 & ~0 \cr
{}~0 & ~1 & ~0 & ~0 \cr
{}~0 & ~0 & ~1 & ~0 \cr
{}~0 & ~0 & ~0 & ~1 \cr
-1 & -1 & -1 & -1 \cr
}\right )
\end{equation}
The charge of the graviphoton in all 5 holes equals the mass and vanishes. The
mass
formula is
\begin{equation}
m_a = {\sum_{I=1}^{I=4} Z^I_{\infty} q^I{}_a \over (1-|Z_\infty|^2)} \ .
\end{equation}
The massless solution is
\begin{equation}\label{5crtical}
 Z^I   =  \sum _{a= 1}^{a=5} {q^I{}_a \over |\vec x - \vec x_a| }\, \ , \qquad
\sum _{a= 1}^{a=5} q^I{}_a=0\ , \qquad I=1,2,3,4.
\end{equation}
\begin{equation}
ds^2 = (1-|Z|^2)^{-1}\, dt^2 -
(1-|Z|^2)\,
d\vec x^2 \ .
\end{equation}
The magnetic charges of 4 vector fields are given by $q^I{}_a$. The scalars are
\begin{eqnarray}\label{5}
Z^1 &=& {1 \over |\vec x - \vec x_1| } - {1 \over |\vec x - \vec x_{5}| } \ ,
\\
Z^2 &=& {1 \over |\vec x - \vec x_2| } - {1 \over |\vec x - \vec x_{5}| } \ ,\\
Z^3 &=& {1 \over |\vec x - \vec x_3| } - {1 \over |\vec x - \vec x_{5}| } \ ,\\
Z^{4} &=& {1 \over |\vec x - \vec x_{4}| }- {1 \over |\vec x - \vec x_{5}| } \
{}.
\end{eqnarray}

After the condensation which is described in the dual picture of interacting
hypermultiplets the final system corresponds to the   5-hole solution  with 4
massive and one massless hole. One possible example of such configuration is:
\begin{equation}
m_1 = m_2=m_3=m_4= {C\over \sqrt {1-4C^2}} \,   \qquad m_5=0 \qquad Z^5_\infty
= - Z^6_\infty=Z^7_\infty=-Z^8_\infty=C \  ,
\end{equation}
with the charge matrix
\begin{equation}\label{2}
\left (\matrix{
{}~1 & ~0 & ~0 & ~0 \cr
{}~0 & -1 & ~0 & ~0 \cr
{}~0 & ~0 & ~1 & ~0 \cr
{}~0 & ~0 & ~0 & -1 \cr
-1 & -1 & -1 & -1 \cr
}\right )
\end{equation}
The new 5 holes are sitting in different places\footnote
{The positions of holes in the original multi-center solutions $x_a$ is
arbitrary. We choose a different set of centers  $\vec {\underline {x}}_{a}$
for the  "after phase transition solution"  to stress the fact that the
positions do not have to coincide with the original ones.}
\begin{equation}\label{15}
 Z^I   = Z^I_\infty +  \sum _{a= 1}^{a=5} {q^I{}_a \over |\vec x - \vec
{\underline {x}}_{a}| }\, \ , \qquad Q^5= \sum _{a= 1}^{a=5} q^5{}_a=Q^7= \sum
_{a= 1}^{a=5} q^7{}_a=0 \ ,
\end{equation}

\begin{equation}
Q^6 = \sum _{a= 1}^{a=5} q^6{}_a= -2\ ,  \qquad  Q^8 =\sum _{a= 1}^{a=5}
q^8{}_a=-2 \ .
\end{equation}
The metric is
\begin{equation}\label{39}
ds^2 = \left({1-|Z|^2 \over  1-|Z_\infty |^2}\right)^{-1}\, dt^2 -
 \left({1-|Z|^2 \over  1-|Z_\infty |^2}\right) \,
d\vec x^2 \ ,   \qquad I=5,6,7,8 \ .
\end{equation}
\begin{eqnarray}\label{5}
Z^5 &=& +C + {1 \over |\vec x -\vec {\underline {x}}_{1}| } - {1 \over |\vec x
- \vec {\underline {x}}_{5}| }\ ,\\
Z^6 &= & -C - {1 \over |\vec x - \vec {\underline {x}}_{2}| } - {1 \over |\vec
x - \vec {\underline {x}}_{5}| }\ ,\\
Z^7 &=& +C+ {1 \over |\vec x - \vec {\underline {x}}_{3}| } - {1 \over |\vec x
- \vec {\underline {x}}_{5}| }\ ,\\
Z^8 &=& -C - {1 \over |\vec x - \vec {\underline {x}}_{4}| }- {1 \over |\vec x
- \vec {\underline {x}}_{5}| }\ .
\end{eqnarray}
There are many other solutions with 4 massive and 1 massless holes but with the
different total charge in various gauge groups.
All  of these configurations have   the limit  when the scalars in the vector
multiplets at infinity vanish $C\rightarrow 0$ and all five holes become
massless. However, these solutions still carry a non-vanishing total charge in
some of the gauge groups. In the example above, the charge in the 6-th and 8-th
direction is not vanishing. If we would try to study this system as we did with
the previous one we would find  that
\begin{equation}
<D^{\alpha \beta \,6}> = \sum_{a=1}^{5} q^6{}_a <( h^{* a \alpha} h^{a\beta} +
 h^{* a \beta} h^{a \alpha})> = -2 \, ( v^{*  \alpha} v^{\beta} +
 v^{*  \beta} v^{ \alpha})) \neq 0 .
\end{equation}
Thus $v\neq 0$ is not a ground state of these theory.
Thus the condensation consistent with N=2 supersymmetric potential of
hypermultiplets only
occurs when we  start with  an $n$-hole solution  {\it with the total mass and
all charges vanishing}. For this configuration one can write down the potential
for the hypermultiplets with the global $U(2)$ symmetry with flat directions
along which the holes condense. This symmetry is broken explicitly either when
the mass terms for the hypermultipets is added to the action or spontaneously,
as considered in this paper. After the spontaneous generation of the condensate
(vacuum expectation value of the scalar part of the hypermultiplets in the
effective theory) the new state may be described as an $n$-hole solution of the
same theory but with the different values of masses and charges. The values of
the masses  are defined by  the condensate.

\section{Discussion}
The main result of this work is the realization of the fact that supersymmetric
multi-hole solutions with the vanishing mass and total charges in all gauge
groups do
exist. They are massless and neutral, still it is not a flat space: the
solution has an arbitrary number of centers  with a particular distribution of
charges.
 Such configurations plays the central role   in the picture of  the black hole
condensation.  This picture reflects nicely various properties of unbroken
supersymmetry, in particular, the subtleties of the supersymmetric balance of
forces in different multiplets.

We may qualify the condensation of holes as a process which drives
the  transition from one multi-hole configuration to another. After the
formation of the condensate in the   neutral system of  $(n+1)$ types of
massless holes, we obtain  a new vacuum state with $n$ types of massive holes
and one massless hole. The original gravitational system has
various solutions of this kind which are also supersymmetric but whose total
mass and charges  are not vanishing anymore.
The initial as well as the final
configurations are both exact multi--hole solutions of the same effective
Lagrangian. However these two sets have a dramatic difference in the masses and
charges. In particular one can show that the distribution of charges in the
final state differs from the original one by specific sign-flip of charges in
some holes and by different asymptotic behavior of  the scalars in the vector
multiplets.
 Therefore both the original as well as the final set of holes are in
equilibrium according to the balance of forces analysis in supersymmetric
systems.
The picture of  transition between two different sets of supersymmetric holes
presented above relies on the effective theory and
may suggest the possible link between various supersymmetric solitons which are
known to exist in supergravities and in string theories. The ultimate
importance of the spontaneous generation of the central charges in
supersymmetric theories reveals itself in this picture: the massive
supersymmetric configurations are created in the process of spontaneous
breaking of symmetries of the theory. The mass parameters of the theory are not
fixed at the fundamental level, but appear  in the theory with the various
choices of the ground states. We have found that the picture of condensation of
N=2 supersymmetric BPS states is related to the spontaneous breaking of a
global $U(2)$ symmetry. The presence of central charges in $N$-extended
supersymmetry is known to break the global $U(N)$ symmetry. One can expect that
the dynamics of the massive BPS states which form short massive multiplets in
general may be understood via spontaneous breaking of the $U(N)$ symmetry which
is present  in massless theories with global extended $N$ supersymmetries.

 We would like to add, from a somewhat different perspective, that the
existence of flat directions often has  important cosmological implications
such as inflation, Polonyi field problem, etc. It would be very interesting to
study this question in the context of the theory of BPS condensation.  As a
first step one may try to obtain solutions describing massless multi center BPS
states in de Sitter space, like it was done in \cite{KT} for extreme
Reissner-Nordstr\"{o}m  black holes. Since   massless black holes do not modify
metric of space-time far away from them, it may happen that they will not
affect de Sitter expansion, and, {\it vice versa},   flat directions which we
discussed above, will remain flat in de Sitter space. If this is the case, then
one may expect that during inflation quantum  fluctuations will move the  BPS
field $h^{a\alpha}$ along all possible flat  directions in different causally
disconnected parts of the universe. This would divide the universe after
inflation into exponentially large domains corresponding to different stringy
vacua.  One should note that it is not so easy to obtain   inflation in string
theory. But here again the existence of flat directions of a new type may occur
to be very useful. The best way of unifying string theory and inflation would
be to consider inflation along flat directions. If some of the BPS flat
directions become not exactly flat either due to expansion of the universe or
because of the supersymmetry breaking, one may try to investigate the
possibility that inflation occurs during the rolling of the black hole
condensate towards the minimum of its effective potential. This is admittedly a
very speculative possibility, but it sounds so interesting that it would be
hard not to mention it here.

\section*{Acknowledgements}

We are grateful  to   L. Susskind   for a stimulating  discussion. This
work was  supported by  NSF grant PHY-8612280.


\newpage

\begin{thebibliography}{1000}
\bibitem{Klaus2} K. Behrndt,  ``About a class of exact string backgrounds,''
Humboldt--Universit\"at preprint HUB-EP-9516 (1995), hep--th/9506106.
\bibitem{K} R. Kallosh, ``Duality symmetric quantization of
superstring,'' Stanford University preprint SU-ITP-95-12 (1995),
hep-th/9506113, to be published in Phys. Rev. D.
\bibitem{KL}  R. Kallosh and  A. Linde, ``Exact Supersymmetric Massive and
Massless White Holes,''   preprint SU-ITP-95-14 (1995),
hep-th/9507022, to be published in Phys. Rev. D.
\bibitem{CY} M. Cveti\v c and D. Youm, Phys. Lett. {\bf B359},  87 (1995).
\bibitem{FKS} S. Ferrara, R. Kallosh, and A. Strominger, ``N=2 extremal black
holes," preprint CERN-TH-95-211 (1995),
hep-th/9508072, to be published in Phys. Rev. D.
\bibitem{Town} P. Townsend, ``Supergravity solitons and non-perturbative
superstrings," preprint DAMTP-R-95-52 (1995), hep-th/9510190.
\bibitem {HullT} C. Hull and P. Townsend, Nucl. Phys. {\bf B438},  109 (1995);
Nucl. Phys. {\bf B451},  525 (1995).
\bibitem {W} E. Witten, Nucl. Phys. {\bf B443},  85 (1995).
\bibitem{Strom} A. Strominger, Nucl. Phys. {\bf B451},  96 (1995).
\bibitem{US} R.~Kallosh, A.~Linde, T.~Ort\'{\i}n,
A.~Peet, and A.~van Proeyen,   Phys.~Rev. D {\bf  46},  5278 (1992).
\bibitem{BK} K. Behrndt and  R. Kallosh, ``O(6,22) BPS configurations of the
heterotic string,''
 SU-ITP-95-19  (1995),  hep-th/9509102.
\bibitem {Sen} A. Sen, Int. J. Mod. Phys.  {\bf A8}, 5079 (1993);
Nucl. Phys. {\bf B440}, 421 (1995).
\bibitem{GMS} B.R. Greene, D.R. Morrison, and A. Strominger, Nucl. Phys. {\bf
B451},  109 (1995).
\bibitem{KT} D. Kastor and J. Traschen, Phys. Rev. D{\bf 47}, 5370 (1993).



\end{thebibliography}
\end{document}